\documentclass[12pt,a4paper]{article}
\usepackage[english]{babel}
\usepackage{graphicx}
\newcommand{\alt}{\mathbin{\lower 3pt\hbox
   {$\rlap{\raise 5pt\hbox{$\char'074$}}\mathchar"7218$}}}
\newcommand{\agt}{\mathbin{\lower 3pt\hbox
   {$\rlap{\raise 5pt\hbox{$\char'076$}}\mathchar"7218$}}}

\begin{document}
\setcounter{footnote}{0}
\setcounter{equation}{0}
\setcounter{figure}{0}
\setcounter{table}{0}
\vspace*{5mm}

\begin{center}
{\large\bf High Orders of Perturbation Theory:  \\
Are Renormalons  Significant? }

\vspace{4mm}
I. M. Suslov \\
P.L.Kapitza Institute for Physical Problems,
\\ 117337 Moscow, Russia \\
\vspace{6mm}

\begin{minipage}{135mm}
{\rm {\quad}
According to Lipatov, high
orders of perturbation theory are determined by the saddle-point
configurations (instantons) of the corresponding functional
integrals.  According to 't Hooft, some
individual large diagrams, renormalons,  are also
significant and they
are not contained in the Lipatov contribution.
The history of the conception of renormalons is
presented, and the arguments in favor of and against their
significance are discussed.  The analytic properties of the Borel
transforms of functional integrals, Green functions, vertex
parts, and scaling functions are investigated in the case
of  $\varphi^4$
theory.  Their analyticity in a complex plane with a cut from the
first instanton singularity to infinity (the
Le~Guillou\,--\,Zinn-Justin hypothesis) is proved.  It rules out
the existence of the renormalon singularities pointed out by
't Hooft and demonstrates the nonconstructiveness of the
conception of renormalons as a whole.  The results can be
interpreted as an indication of the internal consistency
of  $\varphi^4$
theory.
}
\end{minipage} \end{center}

\newpage
\vspace{6mm}
\begin{center}
{\bf 1. INTRODUCTION} \\
\end{center}

Many problems in theoretical physics can be reduced to a
calculation of functional integrals of the type
$$
I = \int D\varphi\exp(-S_0\{\varphi\} - gS_{\rm int}\{\varphi\}),
\eqno(1)
$$
whose expansion in the coupling constant $g$ gives an ordinary
perturbation theory.  In 1977 Lipatov \cite{1} proposed a method
for calculating the high-order expansion coefficients of the
integrals (1) on the basis of the following simple idea.  If the
function $F(g)$ is expanded into a series
$$
F(g) = \sum\limits_{N=0}^\infty F_Ng^N,
$$
the $N$th expansion coefficient can be calculated from the
formula
$$
F_N = \oint\limits_C \frac{dg}{2\pi i}\frac{F(g)}{g^{N+1}},
\eqno(2)
$$
where the contour $C$ encloses the point $g=0$ in the complex
plane.  Taking the integral (1) as $F(g)$, we obtain
$$
I_{N-1} = \frac{1}{2\pi i}\int dg \int D\varphi\exp(-S_0\{\varphi\} -
 gS_{\rm int}\{\varphi\}
-N\ln g),
\eqno(3)
$$
and the appearance of an exponential  with a large
exponent  indicates that the saddle-point method
can be applicable at large $N$.  Lipatov's idea is to
seek the saddle point in
(3) with respect to $g$ and $\varphi$ simultaneously:
such saddle-point exists for
all the cases of interest and is realized on a spatially
localized function, which has been termed an
instanton.  Moreover, the conditions for applicability of the
saddle-point method are satisfied at large $N$.

The Lipatov technique, which was originally applied to
scalar theories, such as $\varphi^4$  theory
$$
S_0\{\varphi\} + gS_{\rm int}\{\varphi\}= \int d^dx
 \left\{\frac{1}{2}(\nabla \varphi)^2 +
\frac{1}{2}m^2\varphi^2 + \frac{1}{4}g\varphi^4 \right\},
\eqno(4)
$$
was subsequently generalized to vector fields \cite{2}, scalar
electrodynamics \cite{3,4}, Yang--Mills fields \cite{5}, fermion
 fields \cite{6}, etc.
(see the collection of articles in Ref. \cite{7}).  The
ultimate goal was to apply it to theories of practical interest,
viz., quantum electrodynamics \cite{8,9} and quantum
chromodynamics (QCD) \cite{10,11}.  As was pointed out already in
Lipatov's first paper \cite{1}, the knowledge of the first few
coefficients and their asymptotic behavior permits approximate
reconstruction of the Gell-Mann--Low function, opening up a
direct route to the solutions of the problem of confinement and
electrodynamics at short distances.

However, a conception which raised some doubts regarding the
Lipatov technique was conceived already in 1977.  It was
initiated by a paper by Lautrup \cite{12}, which contained the
following curious observation.  The typical result of
calculations based on the Lipatov technique has a functional form
$$
I_N = ca^N\Gamma(N + b) \approx ca^NN^{b-1}N!
\eqno(5)
$$
and the natural interpretation of it is that there is a
factorially large number of diagrams of the same order $(ag)^N$.
However, in the general case such an interpretation is incorrect,
since there are examples of individual $N$th-order diagrams
having a value ${\sim}N!$.  The latter are diagrams
(Fig.\,1) which contain long chains of ``bubbles''.
\begin{figure}
\centerline{\includegraphics[width=5.1 in]{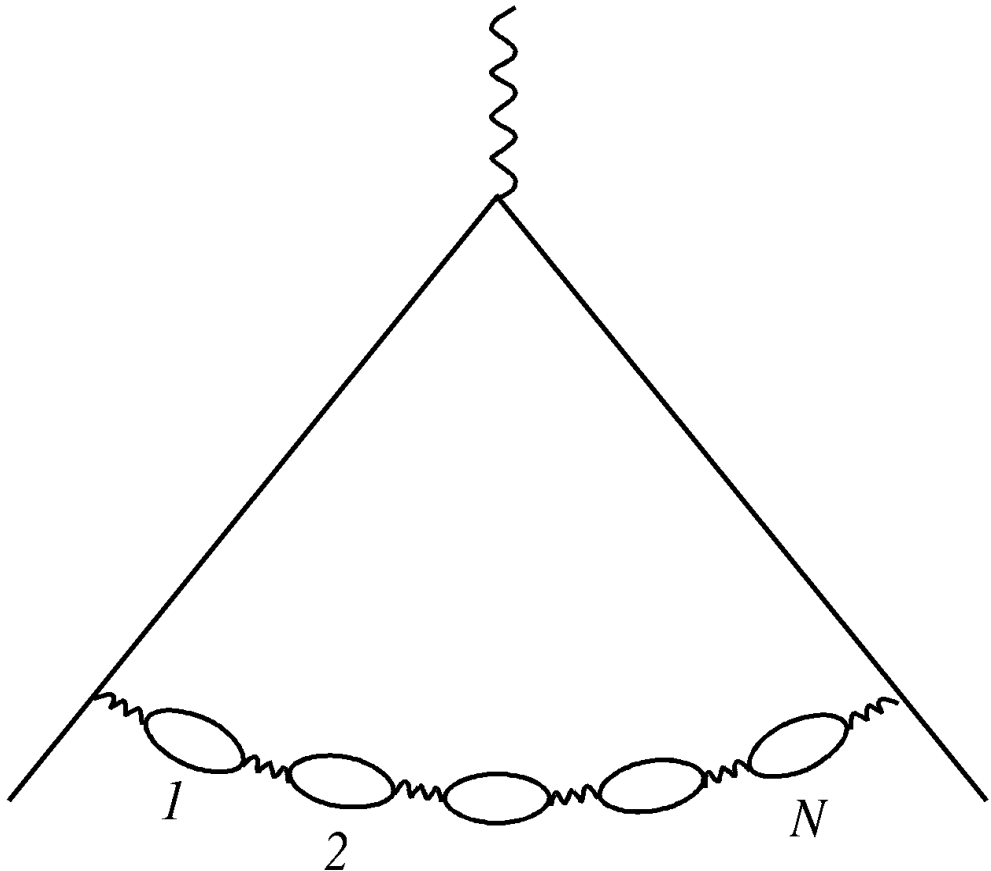}}
\caption{\,\,Example of a diagram for quantum electrodynamics, which
makes a contribution ${\sim}N!$ to the $N$th order of perturbation
theory \cite{12}.} \label{fig1}
\end{figure}
Such factorial contributions of individual diagrams
were termed renormalons, since they appear only in renormalizable
theories.\,\footnote{\,In a broader sense, a renormalizable
theory is one in which the divergences are eliminated by
renormalizing a finite number of parameters.  According to more
precise terminology, such theories are subdivided into
super-renormalizable (renormalizable ``with a surplus'') and
renormalizable in the narrow sense (marginally renormalizable);
the latter, which gave their name to renormalons, are
characterized by the logarithmic situation, which is needed for
the appearance of factorial contributions (see below).
}
Lautrup's example (Fig.\,1) was
related to quantum electrodynamics, but similar diagrams exist in
QCD and four-dimensional $\varphi^4$ theory.

Strictly speaking, nothing followed from Lautrup's observation:
the Lipatov technique is based on a formal calculation of the
functional integral (3) and does not rely in any way on a
statistical analysis of diagrams.  It is natural to expect that
the renormalon contributions have already been taken into account
in the instanton result (5).  In fact, no far-reaching claims
were made in Ref. \cite{12} or in the relevant publications
appearing shortly thereafter \cite{13,14}.

However, the tone of the publications
subsequently changed dramatically.  The reason was
't Hooft's lecture \cite{15}, which was delivered in the same
year, 1977.  The term "renormalon" was used in it for the first
time, and it was asserted that renormalons are not contained in
the instanton contribution (5).  The authors of the subsequent
publications
\cite{16}--\cite{30}
considered 't Hooft's opinion to be self-evident and did not
trouble themselves with argumentation.

A convenient language for discussion, viz., the analytic
properties of Borel transforms, was introduced in 't Hooft's
lecture.  The Borel transformation
$$
F(g) = \sum\limits_{N=0}^\infty F_Ng^N = \sum\limits_{N=0}^\infty
\frac{F_N}{N!}\int\limits_0^\infty  dx\,x^Ne^{-x}g = \int\limits_0^\infty
 dx\,e^{-x}
\sum\limits_{N=0}^\infty\frac{F_N}{N!}(gx)^N,
$$
which factorially improves the convergence,
is widely used in the theory of divergent series
\cite{30}.  It is
convenient to rewrite it in the form
$$
F(g) = \int\limits_0^\infty dx\,e^{-x} B(gx),
\eqno(6)
$$
$$
B(z) = \sum\limits_{N=0}^\infty \frac{F_N}{N!}z^N
\eqno(7)
$$
by introducing the Borel transform $B(z)$ of the function $F(g)$.
The Borel transform for a function with the expansion
coefficients (5)
$$
B_I(z) = \sum\limits_N ca^NN^{b-1}z^N \sim (1-az)^{-b},\ \ za \rightarrow
 1
\eqno(8)
$$
has a singularity at the point $z = 1/a$.

't Hooft arrived at this result in a different way, without
reference to the Lipatov technique.  Rewriting the integral (1)
and the definition of the Borel transform (6) in the form
$$
I = \int D\varphi\exp(-S\{\varphi\}/g),
\eqno(9)
$$
$$
F(g) = \int\limits_0^\infty dx\,e^{-x/g}B(x),
\eqno(10)
$$
which can be accomplished by means of the replacements $\varphi \rightarrow
\varphi/\sqrt{g}$ and $x \rightarrow x/g$ in (4) and
 (6),\,\footnote{\,'t Hooft omitted the factors of the form $g^n$,
since integrals of the type (1) usually appear in the form of a
ratio and such factors cancel out.
}
  yields the Borel transform of the integral (9):
$$
B_I(z) = \int D\varphi\delta(z - S\{\varphi\}) =
 \oint\limits_{z=S\{\varphi\}}\frac{d\sigma}{|\nabla
S\{\varphi\}|},
\eqno(11)
$$
where the latter integration is carried out over the hypersurface
$z = S\{\varphi\}$.  If an instanton $\varphi_c(x)$, i.e., a classical
solution with a finite action, exists for the integral (9), then
$\delta S\{\varphi_c\} = 0$ and the partial derivatives $\partial
 S/\partial\varphi_i$ with respect
to all the variables $\varphi_i$ comprising $D\varphi$ vanish; therefore,
$\nabla S\{\varphi_c\} = 0$ and the Borel transform (11) has a
singularity at the point
$$
z = S\{\varphi_c\},
\eqno(12)
$$
which coincides with $1/a$.  In addition, there are singularities
at the points $mS\{\varphi_c\}$, which correspond to solutions
containing $m$ infinitely distant instantons.  If it is assumed
that the singularity (12) is closest to the origin of
coordinates, the result (5) of the Lipatov technique is
reproduced.  However, 't Hooft allowed the existence of
singularities differing from those of the instanton type:  in
this case the asymptotics of the expansion coefficients can be
specified by the singularity which is closest to the origin of
coordinates.

't Hooft regarded renormalons as a possible mechanism for the
appearance of the new singularities.  Let us take an arbitrary
diagram for quantum electrodynamics and single out the line of a
virtual photon with the
momentum $k$ (or an interaction line in $\varphi^4$
theory) in it (Fig.\,2,a):  it corresponds to integration
\begin{figure}
\centerline{\includegraphics[width=5.1 in]{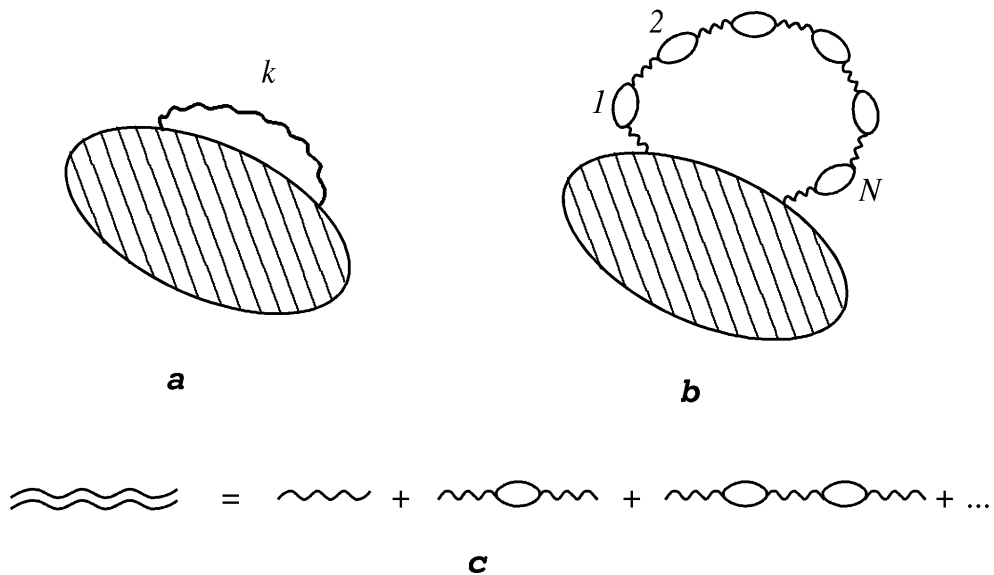}} \caption{
\,\,More
general class of renormalon diagrams} \label{fig2}
\end{figure}
over a region of large momenta of the type
$$
\int\limits_{k>k_0} d^4k\,k^{-2m},
$$
where $m$ is an integer.  If we assume that all the
renormalizations have been performed, the integral converges and
$m \geq 3$.  Inserting a chain of $N$ ``bubbles'' into the photon
line (Fig.\,2,b), we obtain the integral
\,\footnote{\,In quantum electrodynamics and QCD a polarization
loop gives the factor $k^2\ln k^2$, and a photon (gluon)
propagator gives $1/k^2$; in four-dimensional $\varphi^4$ theory a closed
loop corresponds to $\ln k^2$, and an interaction line corresponds
to a constant.  In all cases a chain of $N$ ``bubbles'' corresponds
to $(\ln k^2)^N$.}
$$
\int\limits_{k>k_0} d^4k\,k^{-2m}(\beta_0\ln k^2)^N \sim \beta_0^N=
 \int\limits_0^\infty dt\,t^N
e^{-(m-2)t} \sim \left(\frac{\beta_0}{m-2}\right)^N N!.
\eqno(13)
$$
Borel summation of a sequence of such diagrams gives
singularities at the points
$$
z = \frac{m-2}{\beta_0},\ \ m = 3, 4, 5,\ldots.
\eqno(14)
$$
The constant $\beta_0$ is the first nonvanishing expansion coefficient
of the Gell-Mann--Low function (Sec. 2), and with consideration of
the sign relationships ($S\{\varphi_c\} < 0$, $\beta_0 > 0$) 't Hooft arrived
at the picture of singularities for $\varphi^4$ theory shown in
Fig.\,3,a.
\begin{figure}
\centerline{\includegraphics[width=5.1 in]{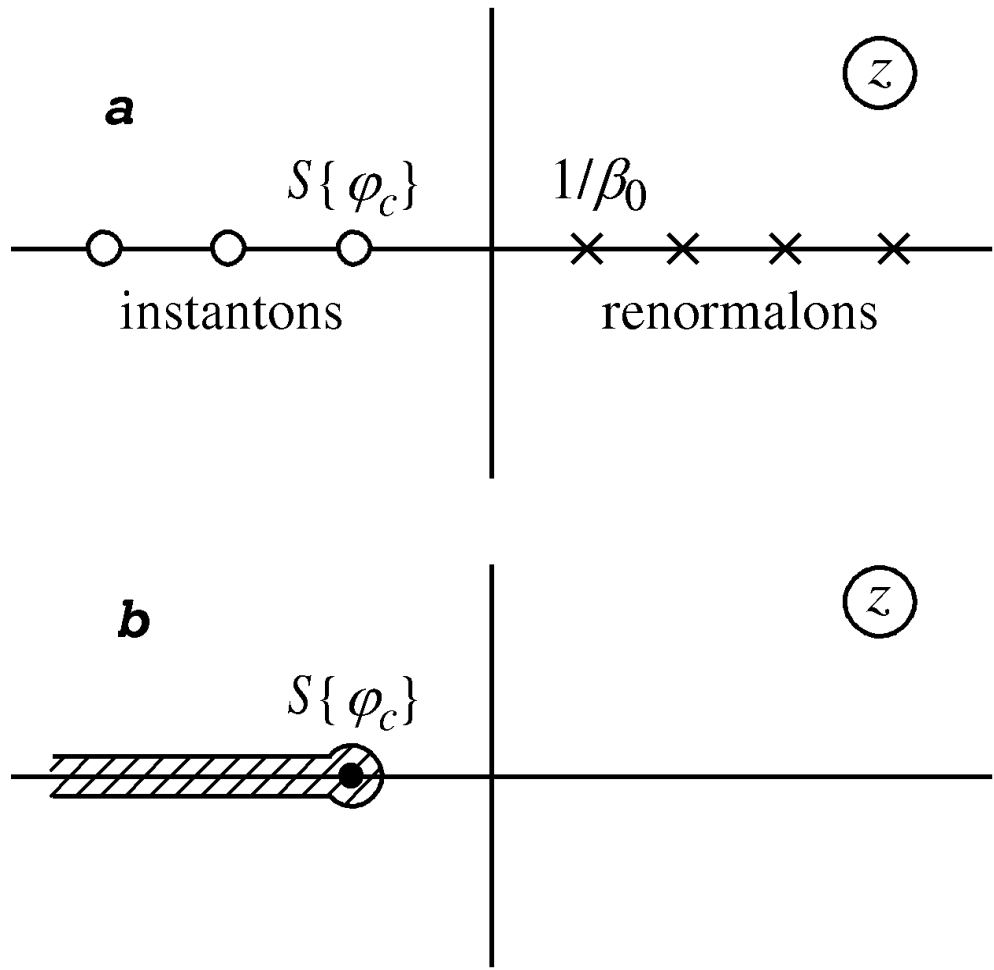}} \caption{
\,\,a) Picture of singularities for $\varphi^4$ theory
given by 't Hooft \cite{15}.
\,\,b) Analyticity region according to the results of the
present work.} \label{fig3}
\end{figure}

It is not difficult to see that 't Hooft's arguments
regarding renormalons leave some fundamental questions unanswered:

Why can significance be attached to individual sequences of
diagrams, which make up an infinitesimal fraction of their total
number when $N \rightarrow \infty$?

How do we know that the renormalons have not already been
taken into account in the instanton contribution (5)?

However, the general setting the question on the
possible contributions of a noninstanton nature to the
asymptotics of the expansion coefficients has a sense:  it
uncovers a gap in the mathematical foundation of the Lipatov
approach.  Indeed, let the function $f(x)$ have a sharp maximum
at the point $x_0$ and a slow tail at large values of $x$
(Fig.\,4,),
\begin{figure}
\centerline{\includegraphics[width=5.1 in]{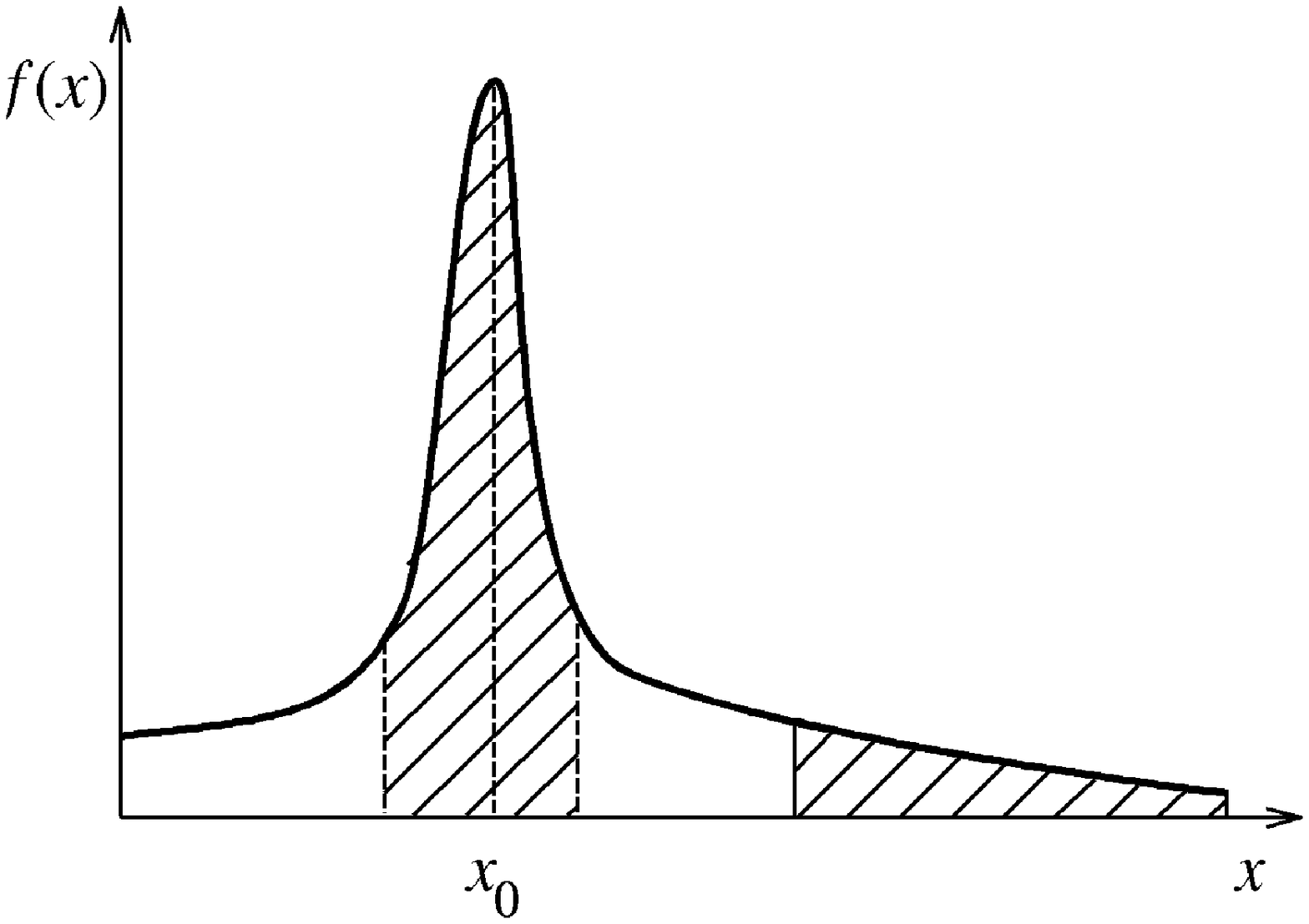}}
\caption{\,\,Example of a function for which the saddle-point method
is formally applicable, but gives a wrong result.} \label{fig4}
\end{figure}
so that the contributions to the integral $\int
f(x)\,dx$ from the vicinity of the maximum and from the tail
region are comparable.  An investigation of the integral for a
saddle point discloses a maximum at $x_0$ and (provided it is
sufficiently sharp) the formal applicability of the saddle-point
method; however, a calculation of the integral in the
saddle-point approximation will be erroneous, since the
contribution of the tail will be lost.  If such tails are present
in the integral (3), the Lipatov technique can be incorrect.
\,\footnote{\,The validity of the saddle-point method can be
substantiated for convergent finite-multiplicity integrals of
functions  $\exp[\lambda F(x)]$ in the limit $\lambda \rightarrow
 \infty$ (Ref.
\cite{31}).  The integral (3) can be brought into the form
indicated, but, generally speaking, it contains both
ultraviolet divergences and divergences associated with
an infinite number of integrations.  The ratio
of two integrals of type (1) must be finite (after the
appropriate renormalizations), but each of them taken
individually can be divergent.
}

The essential lack of nonsaddle-point methods for
calculating functional integrals makes it impossible to straightforwardly
investigate the contribution of possible tails.  But are there
constructive arguments pointing to their existence?  In
principle, such arguments exist, but they have a fairly intuitive
and ambiguous character and do not hold up to criticism when they
are closely examined (Sec. 2).  As a result, the conception of
renormalons has been in a dialectic equilibrium, i.e., it has not
been proved or refuted.  This uncertainty has caused the
interest in high orders of perturbation theory to drop sharply
and Lipatov's program \cite{1} to remain uncompleted.  For
example, a preliminary result for quantum electrodynamics was
obtained back in 1978 \cite{9}, but the parameters $b$ and $c$ in
the asymptotics $I_N = ca^N\Gamma(N/2+b)$ have not yet been
calculated.  Moreover, the first result for QCD appeared in 1991
(Ref. \cite{10}) and was recently revised \cite{11}, but it is
still unsatisfactory (Sec. 4), although the foundation for such
calculations was completely ready in 1980 \cite{5,6}.  Finally,
the attempts to reconstruct the Gell-Mann--Low function have been
restricted to $\varphi^4$ theory \cite{33,34,35}.

A reawakening of interest in asymptotic estimates has
recently been observed, but it has been confined almost
exclusively to the renormalon doctrine
\cite{21}--\cite{30}. In
particular, it is generally accepted (see Zakharov's review
\cite{21})
that renormalons determine the perturbation asymptotics in QCD.
However, the work within the renormalon approach has already
raised some doubts:  the summation of larger sequences of
diagrams leads to dramatic renormalization of the renormalon
contribution and renders the common coefficient in front of them
totally indefinite \cite{30}; in fact, it is impossible to state
that it does not vanish.  On the other hand, the use of the
Lipatov technique has provided significant progress in the theory
of disordered systems \cite{36} and in the theory of
turbulence \cite{37}.

This paper presents a detailed discussion of the existing
arguments in favor of renormalons, which are shown to be unsound
(Sec. 2).  The analytic properties of Borel transforms are
investigated in the example of $\varphi^4$ theory (Sec. 3), and their
analyticity in a complex plane with a cut from the first
instanton singularity to infinity is demonstrated
(Fig.\,3,b).  It rules out the existence of the renormalon
singularities indicated by 't Hooft (Fig.\,3,a) and
demonstrates the nonconstructiveness of the conception of
renormalons as a whole.

An hypothesis that Borel transforms have the analytic
properties indicated was advanced by Le~Guillou and Zinn-Justin
\cite{38}
and underlies one of the most efficient methods for summing
perturbation series, which is based on the use of conformal transformation.
The results obtained below provide the mathematical foundation of this
method.

\vspace{6mm}
\begin{center}
{\bf 2. PROS AND CONS} \\
\end{center}

Let us discuss the arguments in the literature that point to
the existence of noninstanton contributions in the integral (3).

\paragraph*{1.}  There have been numerous semi-intuitive
assertions which reduce to the notion that {\it instantons do not
exhaust all of physics}.  This thesis is correct as long as
it is understood correctly, but in the present case
it is not relevant.

Historically, instantons first appeared when the
saddle-point approximation was employed in the original integral
(1).  It was substantiated only in a narrow region of parameters,
and thus instantons did not, in fact, exhaust all of physics.  In
the Lipatov technique the situation changed, because the
saddle-point approximation is used not in the integral (1), but
in the expression (3) for the expansion coefficients.  Since only
large values of $N$ are considered, only a limited role is
assigned to instantons from the onset; however, the saddle-point
method is now always applicable,
\,\footnote{\,Of course, instantons exist only in a part of the
region of parameters, but this is not a restriction in the
Lipatov technique:  the values of $a$, $b$, and $c$ in the
asymptotics (5) are calculated exactly, and they allow analytic
continuation as functions of the physical parameters.
} and there is a basis to
assume that everything is determined by instantons.

Let us illustrate the foregoing statements in the example of
the Schr\"{o}dinger equation with a random potential $V(x)$:
$$
[\hat{p}^2/2m + V(x)]\Psi(x) = E\Psi(x).
\eqno(15)
$$
At large negative values of $E$ its eigenfunctions (Fig.\,5)
\begin{figure}
\centerline{\includegraphics[width=5.1 in]{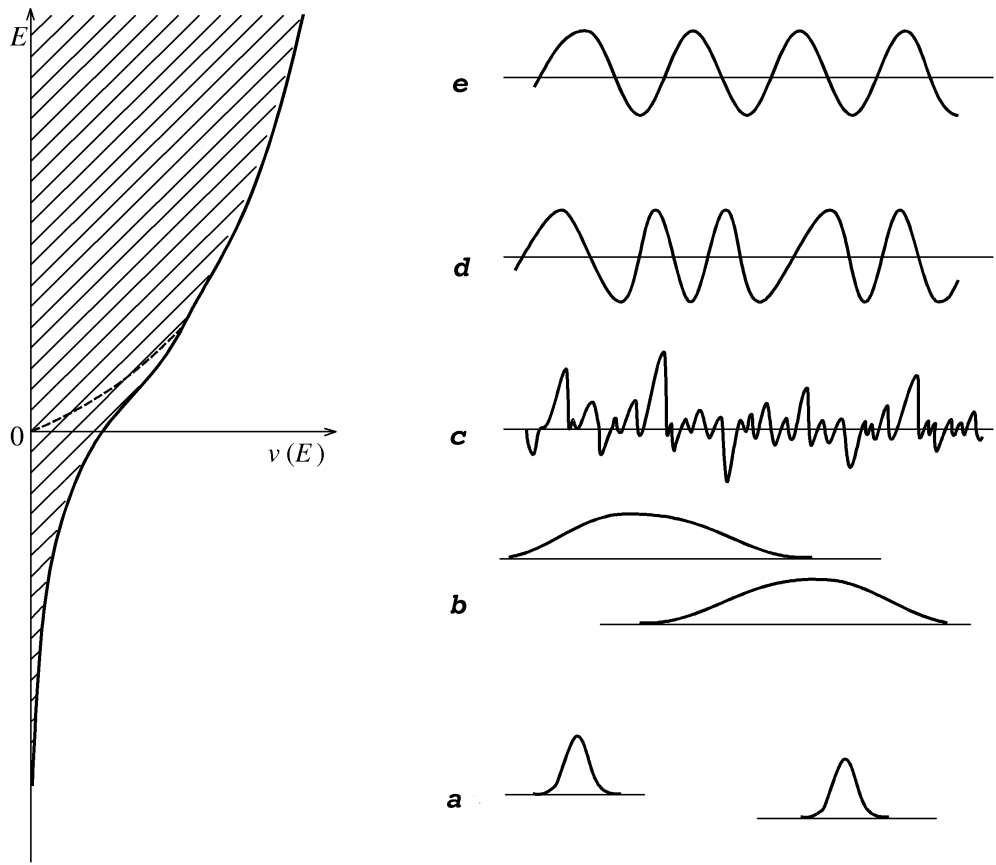}} \caption{
\,\,On
the left --- density of states $\nu(E)$ in the presence (solid
curve) and absence (dashed curve) of a random potential. On the
right --- schematic representations of the eigenfunctions of Eq.
(15).} \label{fig5}
\end{figure}
are localized on the infrequent fluctuations of the random
potential (a and b), at large positive values of $E$ they are
similar to plane waves (d and e), and in the vicinity of the bare
spectrum edge at $E = 0$ they have a highly broken, fractal
character (c).  The problem of investigating Eq. (15) can be
reformulated in the language of an effective field theory, viz.,
$\varphi^4$ theory with the ``incorrect'' sign for $g$ (Refs.
\cite{36} and \cite{39}).  In this case the typical
wave functions of localized states are described by instantons.
The changes in the situation observed as $E$ increases can be
described in the following manner in terms of instantons: at
first instantons have a small radius and a sparse distribution,
i.e., they form an ideal gas (a); then the radius of the
instantons increases, and their density rises, i.e., interactions
between them appear (b); then condensation of the instantons
takes place (c), and an instanton crystal forms (d and e).  Only
the case in Fig.\,5,a corresponds to applicability of the
saddle-point method in an integral of the type (1), and thus the
standard instanton approximation is very poor.

Let us examine this situation from the standpoint of
perturbation theory with respect to the random potential $V(x)$.
An ideal instanton crystal (e) corresponds to a plane wave, i.e.,
the zeroth order of perturbation theory.  In a nonideal crystal
(d) the higher orders have an increased role; in the vicinity of
the bare spectrum edge (c) all the diagrams are of the same order
of magnitude, so that the high and low orders of perturbation
theory are equally significant.  In the region of localized
states (a and b) the dominant role shifts to the high orders:
these states are not manifested in any finite order of
perturbation theory, and discarding the low-order contributions
does not influence their properties in any way.  We see that
Lipatov's conception (high orders are determined by instantons)
fits excellently into the existing physical picture.

Thus, the status of instantons in the integral (1) and the
integral (3) differs significantly.  In our opinion, this
accounts for the position taken by 't Hooft, since he is
a classic on instantons \cite{40} specifically in the original
integral (1).

\paragraph*{2.  Relationship to the logarithmic
situation \cite{15}.}  Renormalons exist only in renormalizable
theories, but not in super-renormalizable theories.  If a theory
is super-renormalizable, an upper bound of the type $a^Ng^N$ can
be obtained for the contribution of an individual diagram, and
the appearance of the factor $N!$ in the asymptotics (5) can be
associated only with a factorially large number of diagrams.
Renormalons and, thus, a new mechanism for the appearance of
factorial contributions appear in renormalizable theories.  It
can be expected that this mechanism is associated with the
formation of the tails in the integral (3) and is not taken into
account in the Lipatov technique.

In this argument everything except the last conclusion is
correct.  We can illustrate this in the case of $\varphi^4$ theory,
which is renormalizable for $d=4$ and super-renormalizable for
$d<4$.  Among the large set of integrations concealed in the
symbol $D\varphi$ in the integral (3), we can single out one for which
the limit $d \rightarrow 4$ is associated with qualitative
changes:  it is the integration over the instanton radius $R$
(Fig.\,6).
\begin{figure}
\centerline{\includegraphics[width=5.1 in]{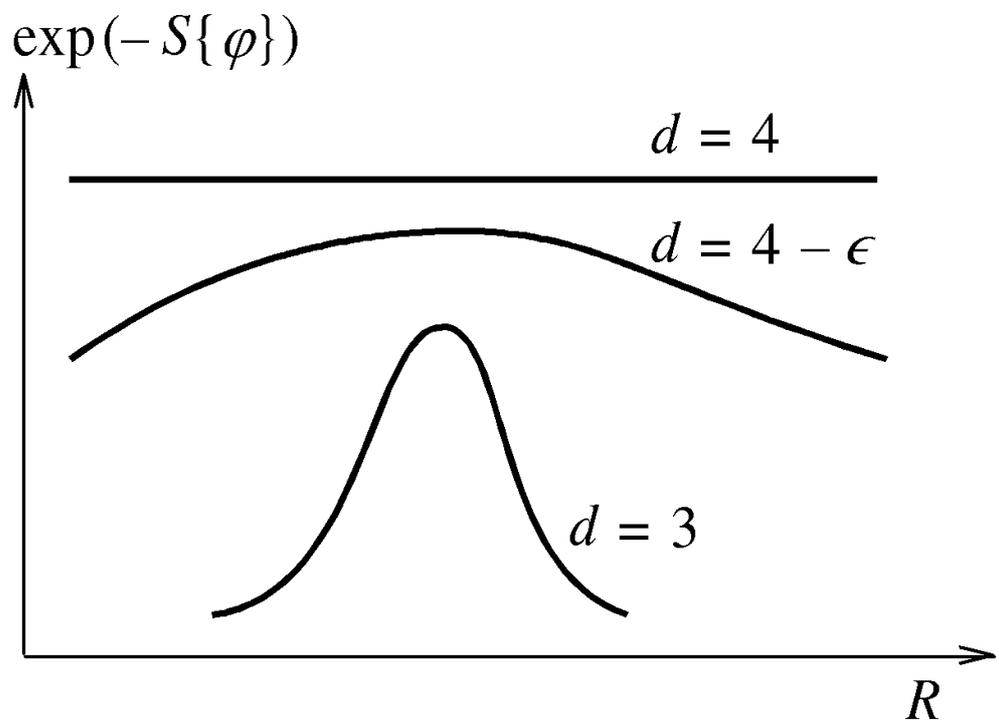}}
\caption{\,\,Dependence of the integrand (1) on the instanton radius
in $d$-dimensional $\varphi^4$ theory.} \label{fig6}
\end{figure}
 For $d$ significantly smaller than 4 (for
example, $d=3$), the integrand $\exp(-S\{\varphi\})$ has a sharp
maximum as a function of $R$ and allows saddle-point integration;
when $d = 4-\epsilon$ the maximum becomes gently sloping, and
when $d=4$ the instanton action $S\{\varphi_c\}$ does not depend on $R$
at all.  In the latter case the integral diverges, leading to the
logarithmic situation.  We see (see the curve for $d=4-\epsilon$
in Fig.\,6) that the ``activation'' of renormalon
contributions is, in fact, related to the appearance of slow
tails in the integral (3), but these tails are taken into account
in the Lipatov technique \cite{36}.

One can be puzzled by a following question.  If the Lipatov technique is
based on a
saddle-point method, then how can it cover the definitely
nonsaddle-point situation for $d=4-\epsilon$?  The fact is that
the saddle point in a functional integral practically never
reduces to a simple maximum achieved at a single point:  the
maximum is degenerate in a certain space of finite
dimensionality.  Accordingly, a finite number of integrations
should be performed exactly, rather than in the saddle-point
approximation.  However, if the integration is performed exactly
over a certain variable (for example, $R$), it is of no
significance whether the degeneracy is exact ($d=4$) or
approximate ($d=4-\epsilon$).  Nevertheless, in the latter case
technical difficulties arise, and the corresponding methods
(constrained instantons \cite{41,42}) have been poorly developed
hitherto \cite{36}.

It is thus clear that even in cases where slow tails
actually appear in the integral (3), the Lipatov procedure is
sufficiently flexible and contains broad possibilities for
dealing with them.

\paragraph*{3.  The limit $n \rightarrow \infty$.}  There is an opinion
that the significance of renormalon contributions can easily be
proved by treating the $n$-component $\varphi^4$ theory in the limit $n
\rightarrow \infty$ (and the analogous models in QCD and
 electrodynamics) \cite{16}:
the factor $n$ corresponds to a closed loop, and renormalon
graphs containing the maximum possible number of loops are
singled out by the large parameter $n$.  Although
diagrams of the same order, but with a smaller number of loops,
can make comparable contributions at large $N$ due to
the combinatorial factors, they have a slower dependence on
$n$; therefore, the renormalons cannot be cancelled identically.

This argument is valid in any finite order of
$1/n$.  However, a detailed investigation of the structure of the
$1/n$-expansion \cite{18,19} reveals the presence of numerous
cancellations, and although the situation cannot be totally
elucidated, the question is not resolved on the level of simple
arguments of the type indicated.

It is not difficult to identify the crux of the problem
here.  As an example, let us consider the self-energy $\Sigma(p, m)$
of $\varphi^4$ theory; it is clear from a diagrammatic analysis for
$m=0$ and values of the momentum $p$ close to the ultraviolet cutoff
$\Lambda$ that the $(N+1)$th expansion coefficient for $\Sigma(p,
0)-\Sigma(0, 0)$ has the form of a polynomial in $n$
$$
p^2\{A_N(N)n^N + A_{N-1}(N)n^{N-1} + \ldots + A_1(N)n + A_0(N)\},
\eqno(16)
$$
in which the coefficient $A_N(N)$ is specified by renormalon
graphs:
$$
A_N(N) = {\rm const}\cdot\left(-\frac{1}{16\pi^2}\right)^NN!.
\eqno(17)
$$
If it is assumed that the renormalon graphs are contained in
the instanton contribution, the expression (16) should transform
into the Lipatov asymptotics at large $N$ [see Eq. (130) in Ref.
\cite{43} for $M=1$ and $p \approx \Lambda$]:
$$
p^2\alpha\beta^nN^{(n+6)/2}\left(-\frac{1}{8\pi^2}\right)^NN!
\left[\Gamma\left(\frac{n+2}{2}\right)\right]^{-1} \int\limits_0^\infty
dy\,y^{(n+5)/3}K_1(y)^2,
\eqno(18)
$$
where $\alpha, \beta \sim 1$, and $K_1(x)$ is the McDonald function.  It is
easily seen that an equality between (16) and (18) is impossible
when $n \rightarrow \infty$.  This is a manifestation of the
 ``noncancelability''
of renormalons.

However, the usual condition for applicability of the
Lipatov technique, $N\gg1$ at large $n$ is, generally speaking,
replaced by a more rigid condition, for example, $N\gg n$, and $n$
then has a bound of the type
$$
n \alt n_0(N),
\eqno(19)
$$
which precludes going to the limit $n \rightarrow \infty$.  If it is taken
 into
account that the Lipatov asymptotics has limited accuracy
(${\sim}1/N$ in relative units), the correct formulation of the
question is as follows.  Can we construct an interpolation
polynomial of type (16) with the high-order coefficient (17)
which would approximate the function (18) within an assigned
accuracy in the interval $0 \leq n \leq n_0(N)$, where $n_0(N) \rightarrow
 \infty$
as $N \rightarrow \infty$?  The answer to this question is positive (see the
Appendix); therefore, the assumption that the renormalon graphs
are contained in the instanton contribution does not
lead to contradictions.

\paragraph*{4.  Relationship to a Landau pole \cite{16,19}.}  It
is easy to see (Fig.\,2,c) that the summation of a sequence of
renormalon diagrams  corresponds to ``dressing'' the interaction.
The relationship between the renormalized charge $g$ and the bare
charge $g_0$ is then given by the familiar
expression \cite{44,45,46}
$$
g_0 = \frac{g}{1-\beta_0g\ln(\Lambda^2/m^2)},
\eqno(20)
$$
which contains a pole at the point
$$
\Lambda_c^2 = m^2e^{1/\beta_0g}.
\eqno(21)
$$
Under the literal understanding of this pole in the spirit of
the early papers by Landau and Pomeranchuk \cite{17}, a simple
physical interpretation can be given to renormalon singularities
(Refs.  \cite{16} and \cite{19}).\,\footnote{\,It is assumed
below that $\beta_0 > 0$.  For asymptotically free theories, in
which $\beta_0 < 0$, similar arguments are valid in regard to
so-called infrared renormalons.  The latter are obtained from
integrals of the type (13) with $m = 1, 0, -1, -2, \ldots$ in the
region of small momenta.     }

The dependence of the perturbation series on the cutoff
parameter $\Lambda$ has the structure
$$
c_{-1}\Lambda^2 + c_0\ln \Lambda^2 + c_1\Lambda^{-2} + c_2\Lambda^{-4} +
 \ldots + c_n\Lambda^{-2n} + \ldots.
\eqno(22)
$$

The first two terms are eliminated by a renormalization
procedure, and the rest terms, in principle, remain, but
vanish in the limit $\Lambda \rightarrow \infty$.  Because of the pole in
 (20), values
of $\Lambda$ greater than $\Lambda_c$ are inaccessible in principle, and
unremovable uncertainties of the type
$$
\Lambda_c^{-2n} = m^2e^{-n/\beta_0g}
\eqno(23)
$$
appear in the theory.  Similar uncertainties are generated by
renormalon singularities, whose existence on the positive
semiaxis leads to ambiguity in the choice of the integration contour
in the Borel integral (6).  The contour can be drawn to the right or
the left of the $n$th singularity (Fig.\,7), producing an
\begin{figure}
\centerline{\includegraphics[width=5.1 in]{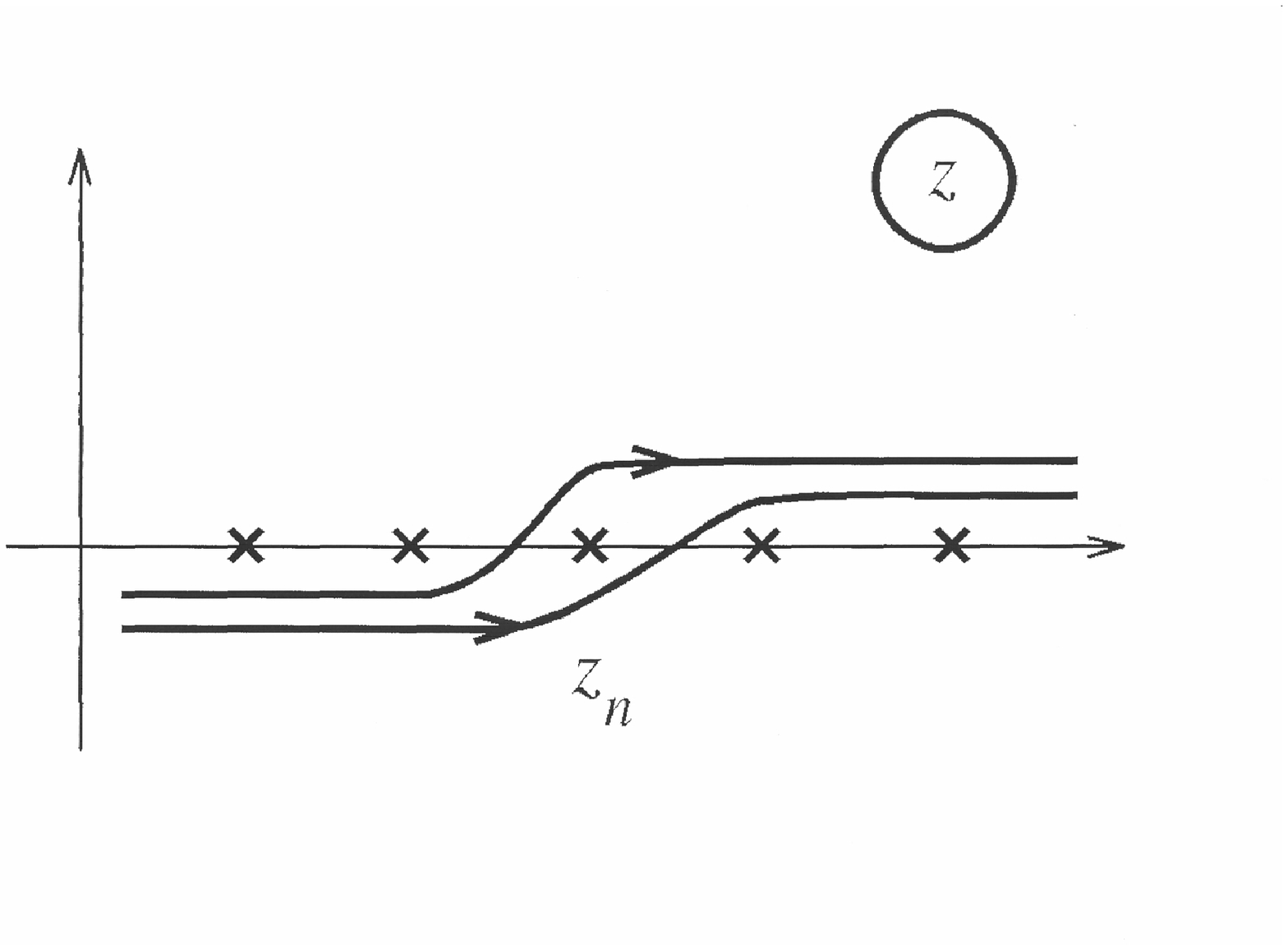}}
\caption{\,\,Ambiguity of the choice of the integration
path in the Borel integral (6) in the presence of
renormalon singularities.}
\label{fig7}
\end{figure}
uncertainty in reconstructing the function from its Borel
transform:
$$
\delta F(g) \sim \oint\limits_{z\approx z_n} dz\,e^{-z/g}B(z) \sim
 e^{-z_n/g},
\eqno(24)
$$
which, with consideration of the equality $z_n = n/\beta_0$ coincides
with (23).

Of course, the literal interpretation of the Landau pole
seems archaic, but after some modification of the argument
presented, a real meaning can be assigned to it.
It is well known \cite{46},
that the dependence of the charge $g$ on the distance
scale $\Lambda^{-1}$ is
given by the equation
$$
\frac{dg}{d\ln \Lambda^2} = \beta(g) = \beta_0g^2 + \beta_1g^3 + \ldots,
\eqno(25)
$$
whose solution depends drastically on the behavior of
the Gell-Mann--Low function $\beta(g)$.  The pole in (20) is
eliminated, if $\beta(g)$ changes sign or behaves as $g^\alpha$ with $\alpha
 <
1$ at large $g$.  If, on the other hand, $\beta(g)$ is positive and
increases as $g^\alpha$ with $\alpha > 1$ when $g \rightarrow \infty$, the
 pole is
preserved, and the theory is internally inconsistent because it
is impossible to determine $g(\Lambda)$ for all $\Lambda$ (Ref.
\cite{46}).  In the latter case the position of the pole is
given by the equation
$$
\ln\frac{\Lambda_c^2}{m^2} =  \int\limits_g^\infty \frac{dx}{\beta(x)},
\eqno(26)
$$
which, for small values of $g$, leads to the result
$$
\Lambda_c^2 = {\rm const}\cdot m^2e^{1/\beta_0g},
\eqno(27)
$$
which differs from (21) only by an insignificant
constant factor.  Thus, the existence of renormalon
singularities seems fairly convincing for internally inconsistent
theories.  Conversely, there is no reason for them in ``good''
theories.\,\footnote{\,In particular, the result (27) is valid when the
expansion (25) is truncated at a finite number of terms, provided
the polynomial obtained is positive.  On these grounds it is easy
to draw the erroneous conclusion that the high-order terms of the
expansion of the $\beta$ function are insignificant.  Parisi's
arguments \cite{16,17} regarding the momentum dependence of Borel
transforms exactly follow this line of reasoning. In fact, the character
of the solution of Parisi's equations
\cite{17} depends significantly on
the behavior of $\beta(g)$ at infinity.  In particular, they are
easily solved for the model function $\beta(g) =
 \beta_0g^2/(1 + \lambda g)$ with
$\lambda \gg 1$ and lead to a result which differs qualitatively from
its one-loop analog.
}

Since the behavior of the function $\beta(g)$ at $g\agt 1$ is
unknown, the presence or absence of renormalon singularities is a
matter of belief.  We stress, however, the following
point.  Factorial contributions of individual diagrams
exist in all field theories in which the expansion of the $\beta$
function (25) begins from the quadratic term: then the
interaction on the $k^{-1}$ scale is described by a formula of the
type (20) with the replacement of $\Lambda$ by $k$, whose expansion
gives $(\beta_0\ln k^2)^N$ in the $N$th order [see (13)].
\,\footnote{\,The concrete sequences of the renormalon diagrams can
differ somewhat in different theories.  For example, in $\varphi^4$
theory the significant diagrams do not reduce to chains of
``bubbles'' (Fig.\,2,c), but form a so-called parquet \cite{48}.
}
To resolve the question of the internal inconsistency of a
theory, one should know all the coefficients in the
expansion (25).  Therefore, it would be incorrect to consider
the formal existence of renormalon contributions as an indication of the
internal inconsistency of a theory.


\vspace{6mm}
\begin{center}
{\bf 3. ANALYTIC PROPERTIES OF THE BOREL TRANSFORMS OF $\varphi^4$
THEORY} \\
\end{center}

\vspace{3mm}
\begin{center}
{\bf 3.1. Expansion of the class of Borel transformations}
 \end{center}

For the ensuing treatment it is convenient to expand the
class of Borel transformations, setting
$$
F(g) = \int\limits_0^\infty dx\,e^{-x}x^{b_0-1}B(gx),
$$
$$
B(g) =  \sum\limits_{N=0}^\infty\frac{F_N}{\Gamma(N + b_0)}g^N
\eqno(28)
$$
with the arbitrary parameter $b_0 >0$, instead of  (6) and
(7).  If $B(g)$ and $\tilde{B}(g)$ are the Borel transforms
corresponding to the parameters $b_0$ and $b_1$ (for definiteness, we
set $b_1 > b_0$), it is not difficult to derive the conversion
formula:
$$
\tilde{B}(g) = \frac{1}{\Gamma(b_1-b_0)}\int\limits_0^\infty dx\,
\frac{x^{b_1-b_0-1}}{(1+x)^{b_1}}B\left(\frac{g}{1+x}\right).
\eqno(29)
$$
We define the analyticity region of $B(g)$ by constructing the
so-called Mittag--Leffler star
\cite{30}, i.e., by drawing cuts from all
singular points to infinity along rays drawn through these
points from the origin of coordinates.  If $g$ lies in the
analyticity region of $B(g)$, the integration path in (29) does
not pass through its singularities and $\tilde{B}(g)$ is also
analytic.  If $g_c$ is a singular point of $B(g)$, the
integration path
in (29) for $g=g_c$ unavoidably passes through $g_c$,
generating a singularity in $\tilde{B}(g)$.  For the interesting
case of power-law singularities we have the correspondence rules
$$
B(g) = A\Gamma(-\beta)\left(\frac{g_c-g}{g_c}\right) \rightarrow
 \tilde{B}(g) = A\Gamma(-\beta
- b_1 + b_0)\left(\frac{g_c-g}{g_c}\right)^{\beta+b_1-b_0}
\eqno(30)
$$
for noninteger $\beta + b_1 - b_0$ and
$$
B(g) = A\Gamma(-\beta)\left(\frac{g_c-g}{g_c}\right)^\beta \rightarrow
 \tilde{B}(g) =
A\frac{(-1)^{n+1}}{n!}\left(\frac{g_c-g}{g_c}\right)^n \ln
\left(\frac{g_c-g}{g_c}\right),
\eqno(31)
$$
if $\beta + b_1 - b_0 = n$ is an integer.

We see that the analyticity region for all the Borel
transforms is identical and that it is sufficient to establish it
for any fixed $b_0$.  The choice $b_0 = 1/2$ is convenient for
investigating functional integrals, since a simple result is
obtained in that case for the Borel transform of an exponential
function:
$$
F(g) = g^{-g} \rightarrow B(g) = \frac{\cos(2\sqrt{g})}{\sqrt{\pi}} =
\frac{1}{2\sqrt{\pi}}\{\exp(2i\sqrt{g} + {\rm c.c.}\},
\eqno(32)
$$
which preserves its exponential form.  This permits writing an
explicit expression for the Borel transform of the functional
integral (1):
$$
B_I(g) = \frac{1}{2\sqrt{\pi}}\int D\varphi\exp(-S_0\{\varphi\})\left[\exp
\left(2i\sqrt{gS_{\rm int}\{\varphi\}}\right) + {\rm c.c.} \right].
\eqno(33)
$$
The integrand is a regular function, and the analyticity region
of $B_I(g)$ is determined by the condition for convergence of the
integral.

\vspace{3mm}
\begin{center}
{\bf 3.2. Analyticity outside the negative semiaxis}
 \end{center}

For simplicity, let us consider scalar $\varphi^4$ theory.
Generalization to the $n$-component case is trivial and reduces
to only some complication of the notation.  We assume that $m^2 >
0$, bearing in mind the subsequent analytic continuation to
arbitrary complex $m^2$.

The integral (33) for $\varphi^4$ theory is defined well for
positive values of $g$, since its convergence is determined by an
exponential function of $-S_0\{\varphi\}$ and is obvious after the
Fourier transformation of $\varphi(x)$:
$$
S_0\{\varphi\}= \frac{1}{2}\int d^dx\, \{(\nabla \varphi)^2 +
 m^2\varphi^2\} =
\frac{1}{2}\sum\limits_k (k^2 + m^2)|\varphi_k|^2.
\eqno(34)
$$
For the analytic continuation to complex $g$ we turn the
integration path in (33), setting
$$
g = \tilde{g}e^{i\Psi},\ \ \varphi = \tilde{\varphi}e^{-i\Psi/4},
\eqno(35)
$$
where $\tilde{g}$ and $\tilde{\varphi}$ are real, and $\tilde{g} > 0$.
Then the integral in (33) takes the form
$$
\int
 D\tilde{\varphi}\exp\left(-S_0\{\tilde{\varphi}\}e^{-i\Psi/2}\right)\left[
\exp
\left(2i\sqrt{\tilde{g}S_{\rm int}\{\tilde{\varphi}\}}\right) + {\rm c.c.}
\right]
\eqno(36)
$$
and converges for $-\pi < \Psi < \pi$.  Thus, the Borel transform is
analytic outside the negative semiaxis.

\vspace{3mm}
\begin{center}
{\bf 3.3. Analyticity within a circle}
 \end{center}

We utilize the formal technique used in Refs. \cite{3}
and \cite{41} and introduce the function
$$
R\{\varphi\}) = \frac{S_0\{\varphi\}^2}{4S_{\rm int}\{\varphi\}}.
\eqno(37)
$$
Then (33) is rewritten in the form
$$
B_I(g) = \frac{1}{2\sqrt{\pi}}\int D\varphi\exp\left(- \left[1 - i
\left(\frac{g}{R\{\varphi\}}\right)^{1/2}  \right]S_0\{\varphi\}\right) +
 {\rm
c.c.},
\eqno(38)
$$
and after the replacement of $R\{\varphi\}$ by the constant $R_0$, it is
analytic within the circle $|g| < R_0$.  Let us now suppose that
$$
R\{\varphi\} \geq R_0
\eqno(39)
$$
for all $\varphi$, i.e., $R_0$ is the exact lower bound of $R\{\varphi\}$.
Setting $g = -|g|e^{i\gamma}$ ($-\pi \leq \gamma \leq \pi$), we have the
inequality
$$
|B_I(g)| \leq \frac{1}{2\sqrt{\pi}}\int D\varphi \left\{\exp \left(-
\left[1 -  \left|\frac{g}{R\{\varphi\}}\right|^{1/2} \cos\frac{\gamma}{2}
\right]S_0\{\varphi\}\right) + \exp(-S_0\{\varphi\})\right\},
\eqno(40)
$$
which ensures convergence of the integral in (33) and,
consequently, its analyticity within the circle $|g| < R_0$.

To find $R_0$, we consider the variational problem of
minimizing $R\{\varphi\}$.  It yields the equation
$$
-\Delta\varphi(x) + m^2\varphi(x) - C\varphi^3(x) = 0.
\eqno(41)
$$
where
$$
C = \frac{S_0\{\varphi\}}{2S_{\rm int}\{\varphi\}},
$$
which, after the replacement $\varphi(x) \rightarrow \varphi(x)/\sqrt{C}$,
 transforms
into the standard equation of an instanton of $\varphi^4$ theory.  Using
it, we can easily show that $R_0 = S\{\varphi_c\}$, which establishes the
required analyticity region (Fig.\,3,b).  Questions
concerning the absence of instantons in massive four-dimensional
theory \cite{47} are discussed in Sec. 3.5.

Apart from the integral (1), some other functional integrals
containing products of the type $\varphi(x_1)\varphi(x_2)\ldots
 \varphi(x_M)$ in the
preexponential factor are of interest. The presence of such
products does not influence the convergence, and all the proofs
performed remain unchanged.

\vspace{3mm}
\begin{center}
{\bf 3.4. Invariance relative to algebraic operations}
 \end{center}

As 't Hooft pointed out \cite{15}, the singularities of Borel
transforms are not shifted when algebraic operations are
performed on the original functions.  This can easily be proved
for a modified definition of the Borel transform (10), which
differs from (6) and (28), since
$$
F(g) = F_0 + F_1g + F_2g^2 + F_3g^3 + \ldots,\ \ B(z) = F_0\delta(z) +
\frac{F_1}{0!}  + \frac{F_2}{1!}z + \frac{F_3}{2!}z^2 + \ldots
\eqno(42)
$$
and $B(z)$ contains a $\delta$-function singularity at zero.  The
transformation of (10) by means of the replacement $g \rightarrow 1/z$
reduces to a Laplace transformation and allows inversion.  It can
be used to express the Borel transform of the product $F_3(g) =
F_1(g)F_2(g)$ in terms of the known Borel transforms of the
factors:
$$
B_3(z) = \int\limits_0^z dz'\, B_1(z')B_2(z-z').
\eqno(43)
$$
It can easily be seen that the $\delta$-function singularity in
$B_3(z)$ corresponds to the definition (42) and that the singular
points for finite $z$ coincide with the singular points of
$B_1(z)$ and $B_2(z)$ (see the analogous reasoning in Sec. 3.1).
In particular, the Borel transform $g^n$ is the function
$z^{n-1}/\Gamma(n)$, which is analytic for integer values of $n$, and
multiplication of the function by $g^n$ does not alter its
analytic properties in the Borel plane.

If $F_2(z) = 1/F_1(z)$, then
$$
\delta(z) = \int\limits_0^z dz'\, B_1(z')B_2(z - z')
\eqno(44)
$$
and the $\delta$-function singularity on the left-hand side cancels
out with the $\delta$-function singularities in $B_1(z)$ and $B_2(z)$.
At finite values of $z$ the right-hand side contains
singularities corresponding to singular points of $B_1(z)$ and
$B_2(z)$, which are absent on the left-hand side  and, therefore,
compensate one another.  This is possible only if $B_2(z)$ has
singularities at the same points as $B_1(z)$.

The proof of the analogous statements for linear operations,
viz., summation, differentiation, integration, etc., is trivial.

The standard definition of the Borel transform (6) is
obtained from (10) and (42) when $F_0 = 0$ after the replacement
$F(g) \rightarrow gF(g)$.  In this case the $\delta$-function singularities
disappear, and the remaining singularities are preserved at the
same points due to the insignificance of the multiplier $g$.  The
definition (6) corresponds to the definition (28) with $b_0 = 1$,
and, by virtue of Sec. 3.1, the analysis performed can be
extended to arbitrary $b_0$.

Since all the quantities entering into the theory, viz. the
Green's functions, vertex parts, etc., can be expressed in terms
of functional integrals with identical analytic properties
(see the end of Sec. 3.3) using a finite number of
algebraic operations, their singular
points in the Borel plane are the same as for the integral (1).

\vspace{3mm}
\begin{center}
{\bf 3.5. Renormalization procedure}
 \end{center}

The absence of ultraviolet divergences was implicitly
assumed above.  In $\varphi^4$ theory this is correct for $d < 2$.  For
$2\leq d\leq 4$  a continual theory without divergences can be
constructed
by introducing counterterms into the Lagrangian
\cite{46,50}.  In
the simple case where only renormalization of the mass is
required ($2\leq d< 4$) the corresponding term in (4) is
rewritten in the form
$$
m_0^2\varphi^2 = (m^2 + \Delta m^2)\varphi^2 = (m^2 + Ag + Bg^2 + Cg^3 +
 \ldots)\varphi^2,
\eqno(45)
$$
where the coefficients $A, B, C, \ldots$ are chosen so as to
cancel the divergences.  When counterterms are present, the
analytic properties of integrals of the type (1) become more
complicated, since the coupling constant appears not only in the
combination $g\varphi^4$, but also in the form of $g\varphi^2$,
 $g^2\varphi^2$, etc.
One of the directions of renormalon-related activity involved
specifically the introduction of additional terms into the
Lagrangian and tracing the renormalon singularities
appearing \cite{16,18,19}.  A question arises in regard to the
cancellation of singularities in the case when the coefficients
in front of the additional terms are selected so as to remove
divergences, for which an unequivocal answer could not be obtained.

A simpler route is to explicitly introduce regularization
and to use renormalization-group equations.  Here we have in mind
the so-called cutoff scheme \cite{51}:  the vertex parts are
calculated perturbatively as functions of the bare charge $g_0$ and the
cutoff parameter $\Lambda$, then scaling
functions which depend only on $g_0$ are obtained, and, finally,
renormalized vertices, which depend on the renormalized charge
$g$, are constructed \cite{50}.  In this case the explicit
introduction of counterterms is not required, but all the details
associated with their presence are taken into account, since the
fundamental possibility of eliminating the divergences is
essentially used to write the renormalization-group equations.

The simplest way of regularization consists of substituting
$\varphi\epsilon(\hat{p})\varphi$, where $\hat{p}$ is the momentum operator,
for the term $(\nabla \varphi)^2$ in (4), which is brought into the form
$-\varphi\Delta\varphi = \varphi\hat{p}^2\varphi$.  If
$$
\epsilon(p) = \epsilon(-p),\ \ \epsilon(p) \geq 0,
\eqno(46)
$$
then both the entire structure of the instanton calculations
\cite{49}
and the proofs presented above are preserved.  The only change
occurs in the equation of the instanton (41), which is brought
into the form
$$
\epsilon(\hat{p})\varphi(x) + m^2\varphi(x) - \varphi^3(x) = 0.
\eqno(47)
$$
When the regularization
$$
\epsilon(p) = p^2 + p^6/\Lambda^4
\eqno(48)
$$
is employed, the dependence of the action $S\{\varphi\}$ on the
instanton radius $R$ in four-dimensional $\varphi^4$ theory has the form
shown in Fig.\,8.
\,\footnote{\,This dependence can easily be obtained by
describing an instanton by two parameters, viz., its radius
and amplitude, and performing a variational estimation of the
action.  In the theory of disordered systems this corresponds to
the optimal-fluctuation method \cite{35}.
}
\begin{figure}
\centerline{\includegraphics[width=5.1 in]{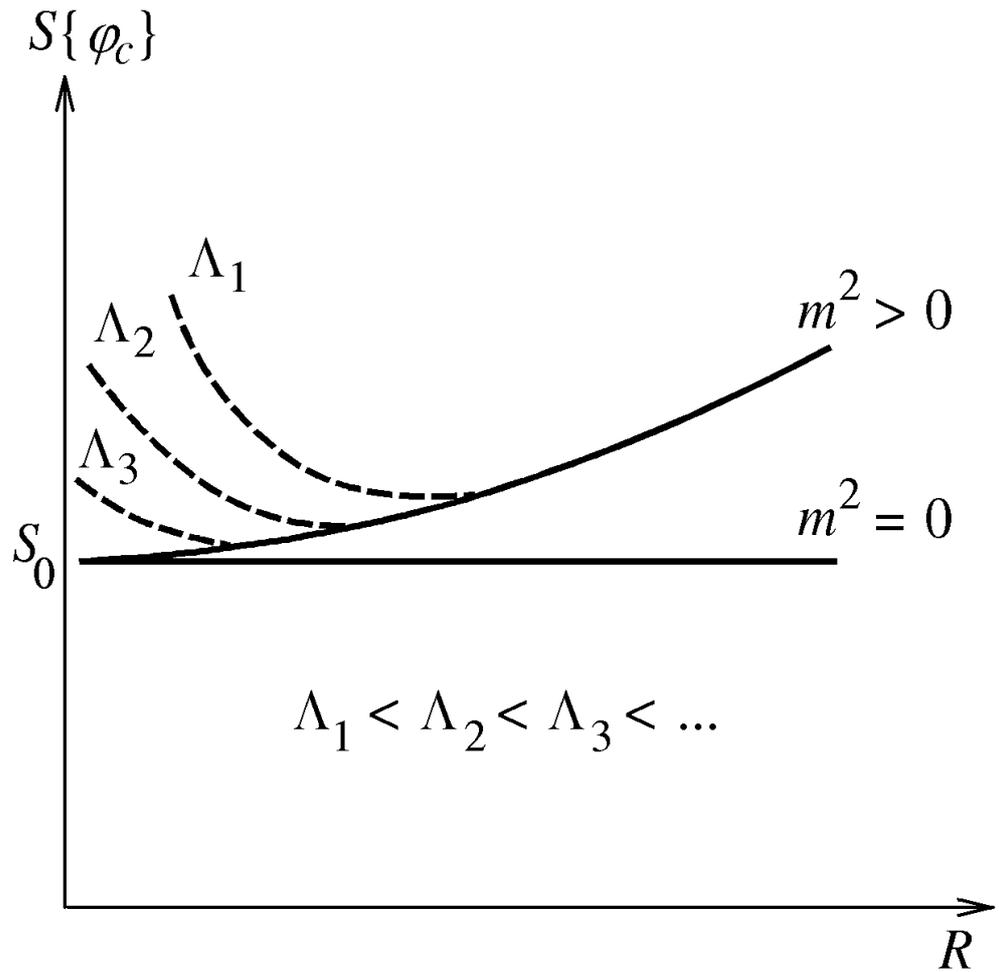}}
\caption{\,\,Dependence of the action $S\{\varphi\}$ on
the instanton
radius $R$ in four-dimensional $\varphi^4$ theory in the absence
of regularization (solid curves) and for a finite cutoff parameter
$\Lambda$ (dashed lines).} \label{fig8}
\end{figure}
If $\Lambda=\infty$, there is
 degeneracy with
respect to the instanton radius in the massless theory
\cite{1}, while
there are no instantons in the massive theory
\cite{49} because of the
monotonic dependence of $S\{\varphi\}$ on $R$.  At finite values of
 $\Lambda$
a minimum appears on the plot of  $S\{\varphi\}$ versus $R$ at $m^2 > 0$
(the dashed curves in Fig.\,8), and instantons appear in the
massive theory.  Their action $S\{\varphi_c\}$ determines the positions
of the singularities in the Borel plane.  For $\Lambda \rightarrow \infty$
 and
arbitrary $m^2 > 0$ the value of $S\{\varphi_c\}$ tends to the instanton
action of the massless theory, and the positions of the
singularities do not depend on $m$.\,\footnote{\,For dimensionalities $2\leq d < 4$ the
influence of $\Lambda$ on the properties of instantons is
insignificant, and the role of renormalizations reduces to
the fact that the instanton equation  contains the renormalized
mass \cite{36}.  The dependence of $S\{\varphi_c\}$ on $m$ is
preserved in this case.  }

The renormalization-group equations (in the Callan--Symanzik
form) are valid for the vertices $\Gamma^{L,N}$ with $N$ free tails and $L$
two-line insertions \cite{50}:
$$
\left[\frac{\partial}{\partial\ln \Lambda^2} +
 \beta(g_0)\frac{\partial}{\partial g_0} + \left(L -
\frac{N}{2}\right)\eta(g_0) - L\eta_2(g_0) \right]\Gamma^{L,N}(g_0, \Lambda)
 = 0.
\eqno(49)
$$
Writing out three such equations with different $L$ and $N$, we
can express the scaling functions $\beta(g_0)$, $\eta(g_0)$, and
 $\eta_2(g_0)$
in terms of the vertices $\Gamma^{L,N}(g_0, \Lambda)$ using algebraic
 operations, which do not shift the positions
of the singularities in the Borel plane.  In the limit $\Lambda \rightarrow
 \infty$,
where Eq. (49) is valid, the dependence of the scaling functions
on $\Lambda$ disappears \cite{50}, and their singularities in the
Borel plane correspond to the massless theory.

The Gell-Mann--Low function $\beta(g_0)$ defines the relationship
between the renormalized charge $g$ and the bare charge $g_0$.
Let the functions $F_0$ and $F_1$ be such that $F_0(g_0)\equiv F_1(g)$.  The
relationship between the corresponding Borel transforms $B_0$ and
$B_1$ [in the sense of the definition (10)] can easily be found
for an infinitesimal charge transformation,  $ g_0 = g +
2\beta(g)\delta\Lambda/\Lambda$ [see (25)]:
$$
B_1(z) = B_0(z) + \frac{2\delta\Lambda}{\Lambda}\int\limits_0^z
 dy\,[B_0(y) +
yB'_0(y)]B_\beta(z-y),
\eqno(50)
$$
where $B_\beta(z)$ is the Borel transform of the function $\beta(g)/g$.
Equation (50) is analogous to Eq. (43); therefore, the analytic
properties do not change in the course of the charge transformation.

The vertices $\Gamma^{L,N}$ diverge as $\Lambda \rightarrow \infty$, but
 they become
finite after separation of the divergent $Z$ factors from them
and the transition from the bare to the renormalized charge.
Since the $Z$ factors are, in turn, expressed in terms of the
vertices $\Gamma^{L,N}$ (Ref. \cite{50}), the renormalized vertices
have the required analytic properties.

The dependence of the scaling functions on the
renormalization scheme is given by the following conversion
formulas \cite{51}:
$$
\tilde{\beta}(q(g)) = \beta(g)\frac{dq(q)}{dg},
$$
$$
\tilde{\eta}(q(g)) = \eta(g) - \beta(g)\frac{d\ln p(q)}{dg},
$$
$$
\tilde{\eta}_2(q(g)) = \eta_2(g) - \beta(g)\frac{d\ln p_2(q)}{dg}.
\eqno(51)
$$
The conversion functions $q(g)$, $p(g)$, and $p_2(g)$
for standard renormalization schemes (subtraction, cutoff
etc.) are expressed in terms of the vertices $\Gamma^{L,N}$.
Therefore, the analytic properties of the scaling functions are
identical in all the schemes.  In the general case the analytic
properties of the conversion functions require additional
investigation.

\vspace{6mm}
\begin{center}
{\bf 4. CONCLUSION} \\
\end{center}

The results in Sec. 3 rule out the existence of renormalon
singularities in $\varphi^4$ theory.  If the arguments in Sec. 2
regarding the relationship to a Landau pole are considered
convincing, $\varphi^4$ theory cannot be internally inconsistent.  The
same conclusion can be drawn on the basis of solid-state
applications:  a reasonable model of a disordered
system reduces exactly to $\varphi^4$ theory \cite{36,39}, and
the internal inconsistency of $\varphi^4$ theory would signify
the impossibility, in principle, of obtaining a mathematical
description of this model.  Therefore, a revision of the results
in Refs. \cite{34} and \cite{35}, in which indications of the
internal inconsistency of $\varphi^4$ theory were obtained on the
basis of an approximate reconstruction of the Gell-Mann--Low
function, is urgently needed.

The results of Sec. 3 refer only to $\varphi^4$ theory and cannot
be extended directly to other field theories; however, along with
the qualitative arguments in Sec. 2, they demonstrate the
nonconstructiveness of the conception of renormalons as a whole.
Therefore, it would be of interest to generalize the method of
proof used in Sec. 3 to other cases.

In quantum chromodynamics (QCD) the renormalon doctrine
presently prevails \cite{20}--\cite{30}.  However, the
 specific
details of QCD in this context have never been stressed.  For
example, 't Hooft \cite{15}, speaking about QCD, gives
explanations within $\varphi^4$ theory and quantum
electrodynamics.  In recent publications \cite{26,30} the term
``naive nonabelianization'' appeared, which essentially means
neglecting the specific details of QCD.  On the other hand, in
QCD there is a special reason for the belief in renormalons,
which has a purely phenomenological character.  The analysis of
experimental data leads to the conclusion that the contribution
of the high orders has a momentum dependence ${\propto}1/q^4$
(Ref. \cite{22}).  This dependence can easily be obtained from
renormalon graphs, but, as is generally assumed, it cannot be
obtained within the instanton method.  The latter is based on the
results in Refs. \cite{10} and \cite{11}, according to which the
instanton contribution is proportional to $1/q^{18}$.  However,
it can easily be seen that a term ${\propto}1/q^4$ appeared in
Refs. \cite{10} and \cite{11}, but contained divergences which
the authors found difficult to eliminate;\,\footnote{\,Such
divergences also appear in $\varphi^4$ theory, and a procedure
for eliminating them is well known \cite{36}. } this term was
``transported'' to the renormalon sector with the reasoning that
it ``contributes to the renormalon singularity \ldots rather than
to the instanton one'' (Ref.  \cite{10}, p. 287).  If there are
no renormalon singularities, this contribution was simply
discarded; therefore, no real calculation of the Lipatov
asymptotics were made for QCD.

This work was stimulated by lengthy discussions with P. G.
Sil'vestrov, whom we thank for opposing the renormalon doctrine,
his critical remarks, and general assistance in acquaintance
with the situation.  We also thank B. L. Ioffe, L. N. Lipatov,
and the participants in the seminars at the Institute of Physical
Problems, the P. N. Lebedev Physics Institute, the Institute of
Theoretical and Experimental Physics, and the St. Petersburg
Nuclear Physics Institute for their interest in this work and
useful discussions.

This work was carried out with financial support from the
INTAS (Grant 96-0580) and the Russian Foundation for Basic
Research (Project 96-02-19527).

\vspace{6mm}
\begin{center}
{\it Appendix}\\
{\bf Construction of interpolation polynomial}
\end{center}

The polynomial of degree $N$ which coincides with the
function $f(x)$ at the points $x_0, x_1, x_2, \ldots, x_N$, is defined
by the Lagrange formula \cite{53}:
$$
P_N(x) =  \sum\limits_{k=0}^N \frac{f(x_k)}{\psi
 '(x_k)}\frac{\psi(x)}{(x - x_k)},
\eqno(A1)
$$
$$
\psi(x) = (x - x_0)(x - x_1)(x - x_2) \ldots (x - x_N),
\eqno(A2)
$$
and the interpolation error is given by the expression
$$
R_N(x) = f(x) - P_N(x) = \frac{f^{(N+1)}(\xi)}{(N+1)!} \psi(x),
\eqno(A3)
$$
where $\xi$ belongs to the interval $(x_0, x_N)$.

The function (18), which is of interest to us, behaves as
$q^n$ at $n \alt N$ with slowly varying $q$, so that $\ln q
\sim \ln N$.  Neglecting these slow variations and omitting the
common multiplier in (16) and (18), we have
$$
f(x) = q^x,\ \ A_N \sim \frac{1}{6^NN^3}.
\eqno(A4)
$$
Taking into account that $|\psi(x)| \leq \Delta^{N+1}$ in the interval $0
\leq x \leq \Delta$, we obtain
$$
|R_N(x)| \leq \frac{(\ln q)^{N+1}q^\Delta}{(N+1)!}\Delta^{N+1},
\eqno(A5)
$$
and the interpolation error is small for
$$
\Delta \alt N/\ln N.
\eqno(A6)
$$

To investigate the dependence of the coefficient $A_N$ on the
positions of the points $x_k$, we set $\psi(x) = \mathop{\rm Re}\psi(x +
i0)$, and calculating $\ln \psi(x + i0)$ using the Euler--MacLaurin
formula, for $\psi(x)$ we obtain the expression
$$
\psi(x) = F(x)\sin G(x).
\eqno(A7)
$$
In particular, for the power-law arrangement of points
$$
x_k = (k/N)^\alpha\Delta,\ \ k = 0, 1, \ldots, N,
\eqno(A8)
$$
at $\alpha \gg1$ we have
$$
F(x) = (-1)^N\sqrt{x(\Delta-x)}\exp
 \left\{N\left[\alpha(x/\Delta)^{1/\alpha} + \ln \Delta -
\alpha\right]\right\},
$$
$$
G(x) = \pi N(x/\Delta)^{1/\alpha}.
\eqno(A9)
$$
For a high-order coefficient of the polynomial (A.1) we obtain
$$
A_N = \sum\limits_{k=0}^N \frac{f(x_k)}{\psi '(x_k)} \sim \exp\{(\alpha
 - \ln \Delta)N\}
\eqno(A10)
$$
(the sum is determined by the term with $k=1$), and for $\alpha \sim
\ln N$ the coefficient $A_N$ can be factorially small or
factorially large, depending on the relationship between $\alpha$ and
$\ln \Delta$, so that the required value of (A4) falls in the range of
variation.  Thus, the required polynomial (16) exists in the
interval $0 \leq n \leq n_0$, where $n_0 \sim N/\ln N$.

Translated by P. Shelnitz

\end{document}